\documentclass[11pt,twoside]{article}
\usepackage{asp2006}
\usepackage{epsf}
\usepackage{psfig}
\usepackage{lscape}
\usepackage{graphicx}

\begin{document}
\title{Spectroscopic monitoring of the Blazar 3C 454.3}
\author{ E. Ben{\'\i}tez\altaffilmark{1}, V. H. Chavushyan\altaffilmark{2}, C. M. Raiteri\altaffilmark{3}, M. Villata\altaffilmark{3},
D. Dultzin\altaffilmark{1}, O. Mart{\'\i}nez\altaffilmark{4}, B. P\'erez-Camargo\altaffilmark{4} and J. Torrealba\altaffilmark{1}}
\altaffiltext{1}{ Instituto de Astronom{\'\i}a, Universidad Nacional Aut\'onoma de M\'exico, Apdo. Postal 70-264,
Ciudad Universitaria, M\'exico DF, M\'exico}
\altaffiltext{2}{Instituto Nacional de Astrof{\'\i}sica, \'Optica y Electr\'onica, Apdo. Postal 51 y 216, 72000, Puebla, Pue., M\'exico}
\altaffiltext{3}{INAF-Osservatorio Astronomico di Torino, I-10025, Pino Torinese, Italy}
\altaffiltext{4}{Benem\'erita Universidad Aut\'onoma de Puebla, Facultad de Ciencias F{\'\i}sico-Matem\'aticas, Apdo. Postal 1152, 72000,
Puebla, Pue., M\'exico}

\begin{abstract}

We performed an optical spectroscopic monitoring of the blazar 3C 454.3 from September 2003 to July 2008. Sixteen optical spectra were obtained during different runs, which constitute
the first spectroscopic monitoring done in the rest-frame UV region (z=0.859). An overall flux variation of the Mg{\small II}($\lambda$2800 \AA) by a factor $\sim3$ was observed, while the corresponding UV continuum ($F_{\rm cont}$ at $\lambda$3000 \AA) changed by a factor $\sim 14$. The Mg{\small II} emission lines respond proportionally to the continuum variations when the source is in a low-activity state. In contrast, near the optical outbursts detected in 2005 and 2007, the Mg{\small II} emission lines showed little response to the continuum flux variations. During the monitored period the UV Fe{\small II} flux changed by a factor $\sim 6$ and correlated with $F_{\rm cont}$ ($r=0.92$). A negative correlation between EW(Mg {\small II}) and $F_{\rm cont}$ was found, i.e.\ the so-called ``Intrinsic Baldwin Effect''.
\end{abstract}

\section{Results}

\begin{figure}[!ht]

\plotone{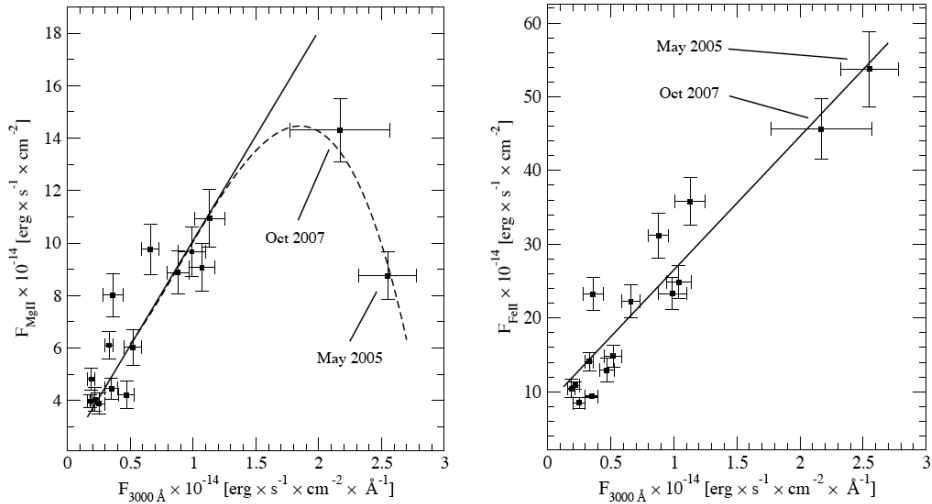}
\caption{{\itshape Left:\/} Mg {\small II} variability behavior during the monitoring period. The solid line represents our linear fit obtained during low-activity states; the broken line shows the deviation from the fit obtained when adding data near the 2005 and 2007 outbursts.
{\itshape Right:\/} The solid line shows the correlation found between the Fe {\small II} and the UV $F_{\rm cont}$ fluxes.}
\end{figure}

In order to model the Mg {\small II} emission resonance doublet at $\lambda\lambda\,\sim$2795,2803 (hereafter Mg {\small II}), we fitted a power law to the continuum emission. In addition, the UV Fe {\small II} template of Vestergaard \& Wilkes (2001) was used for modeling and subtracting the Fe {\small II} contribution in all calibrated and de-reddened spectra. The Mg {\small II} emission line fluxes were obtained after fitting Gaussian profiles. In Figure 1 (left side) we present a plot showing the variation of the $F_{\rm Mg {\small II}}$ vs. $F_{\rm cont}$. When excluding the points near the 2005 and 2007 outbursts, a linear fit was obtained: $F_{\rm Mg {\small II}}=(3.08 \pm 0.67)\,+\, (6.69 \pm 1.05)\,F_{\rm cont}$, with a correlation coefficient $r=0.89$. Including our 2005 and 2007 data, a completely different tendency is observed. On September 3, 2007 (UT) the Mg {\small II} flux is found to be a factor $\sim 1.3$ smaller with respect to the above linear correlation. This tendency is even stronger on May 13, 2005 (UT), where the Mg {\small II} flux is a found to be a factor $\sim 2.3$ smaller. A similar behavior between $F_{\rm Mg {\small II}}$ vs. $F_{\rm cont}$ was found in AO 0235+16 (Raiteri et al.\ 2007a).

Using all spectra we also studied the variability behavior of the UV Fe {\small II} with respect to the continuum changes. We found a linear correlation (see Figure 1, right side): $F_{\rm Fe{\small II}}= (5.12 \pm 2.28) + (23.28 \pm 3.49) \, F_{\rm cont}$, with $r = 0.89$. One of the main results of this study is that the Fe {\small II} flux tends to follow the continuum variations. Although the origin of the UV Fe {\small II} lines in AGN is not completely clarified (Joly et al.\ 2008), it seems possible that part of this emission originates from a region that is close to the base of the relativistic jet, while the Mg {\small II} may be emitted in a more extended region (Goad et al. 1999) and cannot respond immediately to dramatic flux variations. Another possible scenario may be that the relativistic jet continuum swamps almost all the thermal contribution from the Big Blue Bump (see Raiteri et al.\ 2007b) and also from the BLR. Finally, an anticorrelation between EW (Mg {\small II}) and $F_{\rm cont}$ was found using all spectra: EW\,(Mg {\small II})$=\,8.43F_{\rm cont}^{-0.53\pm0.07}$, known as ``Intrinsic Baldwin Effect" (see Osmer \& Shields 1999).

\acknowledgements E.B. acknowledges financial support from grant IN 20308 from PAPIIT-UNAM, and
V.C. and B.P. support from CONACyT-N54480 research grant.

\end{document}